\documentclass[conference, a4paper]{IEEEtran}
\usepackage{graphicx}
\usepackage{soul}
\usepackage[utf8]{inputenc}
\usepackage{lscape}
\usepackage{amsmath}
\usepackage{mathtools}
\usepackage[ruled, vlined, linesnumbered]{algorithm2e}
\usepackage[noend]{algpseudocode}
\usepackage{listings}
\makeatletter
\def\BState{\State\hskip-\ALG@thistlm}
\makeatother
\usepackage{listings}
\usepackage{fancybox}
\usepackage{caption}
\usepackage[font=small,skip=1.5pt]{caption}
\usepackage[ 
natbib=true,
    style=numeric,
    sorting=none]{biblatex}
  \defbibheading{myheading}[References]{%
  \section*{#1}}
\usepackage{balance} 
\begin{document}

\title{Reuse-Aware Cache Partitioning Framework for Data-Sharing Multicore Systems}
 
\author{\IEEEauthorblockN{Soma Niloy Ghosh\IEEEauthorrefmark{1}, Vineet Sahula\IEEEauthorrefmark{2}, Lava Bhargava\IEEEauthorrefmark{3}     }\IEEEauthorblockA{Department of Electronics and Communication Engineering, Malaviya National Institute of Technology, Jaipur, India} Email:\IEEEauthorrefmark{1}2016rec9053@mnit.ac.in, \IEEEauthorrefmark{2}vsahula.ece@mnit.ac.in, \IEEEauthorrefmark{3}lavab@mnit.ac.in}

\maketitle

\begin{abstract}
Multi-core processors  improve performance, but they can create unpredictability owing to shared resources such as caches interfering. Cache partitioning is used to alleviate the Worst-Case Execution Time (WCET) estimation by isolating the shared cache across each thread to reduce interference. It does, however, prohibit data from being transferred between parallel threads running on different cores. In this paper we present (SRCP) a cache replacement mechanism for partitioned caches that is aware of data being shared across threads, prevents shared data from being replicated across partitions and frequently used data from being evicted from caches. Our technique outperforms TA-DRRIP and EHC, which are existing state-of-the-art cache replacement algorithms, by 13.34\% in cache hit-rate and 10.4\% in performance over LRU (least recently used) cache replacement policy.  
\end{abstract} 

\begin{IEEEkeywords} 

WCET, Shared Data, Multi-cores, Cache Partitioning

\end{IEEEkeywords}

\section{Introduction}
Multithreaded applications which share data have not been studied in partitioned caches till date. When applications share information, the benefits of partitioned caches are lessened because duplicate data is put into the partitions, wasting cache space, causing data duplication problems, and degrading cache performance. 

\section{Proposed Approach} \label{sec4}
\subsection{Overview of the SRCP Framework }
 The SRCP framework \cite{ghosh2021srcp} uses \textit{way partitioning} to divide the shared last-level cache (LLC). We performed static partitioning in LLC, a set of cache ways assigned to each core as per equation \ref{eq:1}.
  \begin{equation}
         Partitions \ allotted = \frac{Associativity}{Number\ of\ cores}
         \label{eq:1}
         \end{equation}


 In the SRCP-architecture unlike standard partitioned caches, non-allocated cores can access a partition but can only evict a cache block from its own partition, based on \cite{kedar2017space}. To keep a track of the accesses made to the cache blocks three counters are used, \textit{LC, GCount and AFC} which are dynamically updated and used by the SRCP cache replacement algorithm. 
 The terms used in the cache architecture are as follows:
  \begin{itemize}
     \item \textit{Local Core}: The core allocated to a partition is called local core.
     \item \textit{Global Core}: The cores other than local core are global cores for a partition.
     \item \textit{Local Count (LC)}: A single-bit indicates if a cache block is accessed by the \textit{local core}.
     \item \textit{Access Frequency Count (AFC)}: This is the frequency of accesses made to a block in a cache way by the \textit{local core}. We used \textit{k-bit} counter,where $k = 8$ bits.
   \item \textit{Global Count (GCount)}: This is the number of times  a cache block is accesses by \textit{global cores} within the partition. It's a \textit{n-bit} counter, where $n=log_{2} (no\_of\_cores)$ as the value.
    \end{itemize}
 \subsection{Cache Hit \& Miss Handling}   
 Initially, when a requested cache block is loaded into the LLC the \textit{GCount} is set to \textit{null} and \textit{AFC} counter in \textit{ACT} is set to an intermediate value, \textit{I} ie., 
 \begin{small}
 \begin{equation}\label{eq:2}
 I_{i}=\lceil average(max~\&~min~values~of~AFC_{i}) \rceil,   for ~i^{th}  core. 
 \end{equation}  
 \end{small}
   The \textit{AFC} value of the cache block is increased by one on a hit. Equation \ref{eq:3} specifies the criteria for deciding between frequently utilised and less often used cache lines. A frequently used cache line is loaded into the private cache to increase the cache hits and speed.  Less frequently used data is not loaded into the private cache of a core. 
\begin{small}   
 \begin{equation}\label{eq:3}
   AFC_{i}=\begin{cases}
     freq\_used, & \text{if $AFC_{i} \geq I_{i}$}.\\
     less\_freq\_used, & \text{otherwise}.
   \end{cases}
 \end{equation} 
  \end{small}
  When any \textit{global cores} (other than \textit{local cores}) visit a cache line in a partition, the \textit{GCount} is incremented by one.  Based on the \textit{GCount} given by equation \ref{eq:4}, data in a cache line can be shared or private.
   \begin{small}
  \begin{equation}\label{eq:4}
   Data=\begin{cases}
     Shared, & \text{if $GCount_{i} \geq 1$}.\\
     Private, & \text{otherwise}.
   \end{cases}
 \end{equation}
  \end{small}
 The \textit{AFC \& GCount} values are decreased by one for all the cache blocks in the partition that incurs cache miss when there is a cache miss. As a replacement victim the block with lowest \textit{AFC} value and lowest \textit{GCount} value is picked.  If two or more least frequently used data and least shared data are tied for the selection of a victim cache block, the minimally used block in the recent past by the \textit{local core} is evicted. 
  \subsection{Cache Coherence}
  \textit{Reads and writes} on private data, which are less frequent, are bypassed in our method, as are \textit{writes} on shared data. The dynamic change in application behaviour is taken into account in our method. Because the requested shared cache line will be modified, loading it in the private cache is not allowed. The \textit{write} operation is done directly in the LLC, skipping the L1 cache, resulting in consistent shared data and minimising coherence overheads by retaining only one copy of the shared data.
  
  \subsection{WCET Analysis} 
  Equation \ref{eq:5.1} and \ref{eq:5.2} is used to compute the WCET of shared caches and the proposed framework respectively. The overall latency of a task on a hit is denoted as, $L_{hits}$ while $L_{miss}^{(n-1)}$ denotes latency on a miss, which includes overheads due to threads executing in the other $(n-1)$ cores. $L_{miss}$ is the latency of a task when run solely in the cache partition allotted to it. 
  \begin{small}
   \begin{equation}\label{eq:5.1}
             WCET_{tot} = Cache_{hits} \times L_{hits}~ +  Cache_{miss} \times L_{miss}^{(n-1)}
  \end{equation}
      \begin{equation}\label{eq:5.2}
      WCET_{tot}^{SRCP} = Cache_{hits} \times L_{hits} ~ + Cache_{miss} \times L_{miss}  
  \end{equation}
 \end{small} 
 $ WCET_{tot}>  WCET_{tot}^{SRCP}$ as it includes overheads caused by  shared cache interference due to threads running in remaining $(n-1)$ cores \cite{guo2020inter}. 
\section{Experimental Evaluation} \label{sec5}
\subsection{Experimental Setup}
 The proposed technique was tested using the gem5 full-system simulator. The system parameters used are same as in \cite{ghosh2021srcp}. The multi-threaded Parsec \cite*{bienia2008parsec} and  Splash-2 \cite*{woo1995splash} benchmark suites is used to assess our proposed technique. 
Every application was run for a total of two billion instructions, \& LRU  was utilised as a reference point. The benchmarks are run on four cores with four threads each running on one core, with 16-way associative LLC. Performance is measured in terms of instructions per cycle (IPC).
\subsection{Result \& Analysis}
\begin{figure}[!h]
   	\centering
   	\includegraphics[ 
   	 height=5cm, width=9cm]{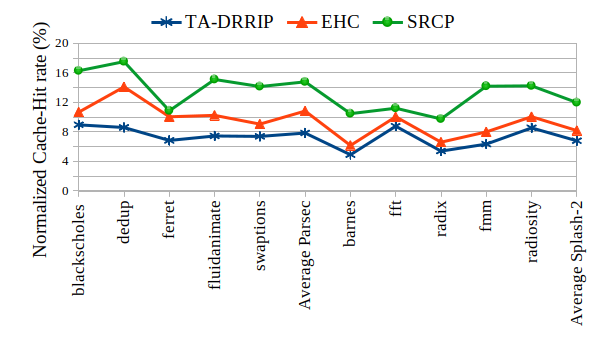} 
   	 	\caption{Increase in LLC hit-rate for Parsec \& Splash-2 benchmarks normalized to LRU. }
   	 
   	 	\label{fig:6a}
   	 \end{figure}

\begin{figure}[!h]
   	\centering
   	\includegraphics[ 
   	height=5cm, width=9cm]{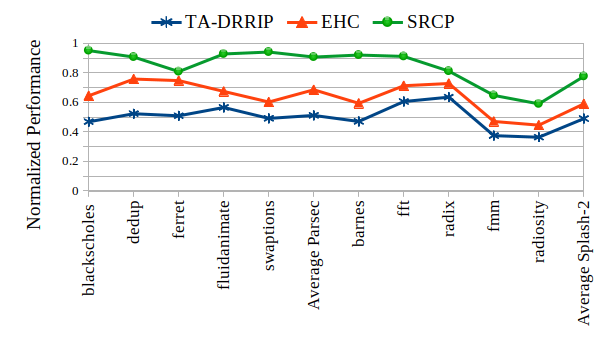} 
   	\caption{Improvement in execution time for Parsec \& Splash-2 benchmarks normalized to LRU. }
   	\label{fig:6b}
  	 \end{figure}

    The improvements in LLC hit-rate and performance are shown in Figures \ref{fig:6a} and \ref{fig:6b}, respectively. The figures shows comparison between our approach ie., SRCP, TA-DRRIP \cite{jaleel2010high} and EHC \cite{vakil2018cache} approaches with LRU as baseline.
    In comparison to TA-DRRIP and EHC, our technique outperforms Splash-2 and Parsec multi-threaded benchmarks. Memory access is quite low for compute heavy applications like \textit{ferret}. Because threads in multi-threaded \textit{ferret} do not coordinate very much, its unlikely that a cache line will be accessed more than once therefore it does not get benefited much with our approach. 
    
    In comparison to LRU, our suggested technique improves cache hit-rate by up to 13.34\%, while EHC and TA-DRRIP boost cache hit-rate by 9.4\% and 7.3\% , respectively. In multi-core CPUs for multi-threaded benchmarks, our technique delivers up to 10.4\% performance gain over LRU, whereas EHC achieves 6.2\% and TA-DRRIP achieves 5\%. 
    
\section{Conclusion} \label{sec6}
This paper presents a simulation model and it is found that our partitioned cache framework is helpful for multi-threaded applications, since it avoids duplication of shared data across cache partitions while also avoiding eviction of shared data. The simulation model uses the already existing model in the gem5 simulator and extends it to add the SRCP features.
\balance
\printbibliography[heading=myheading]
\balance
\end{document}